\documentclass[useAMS]{emulateapj}

\usepackage{epsfig}
\usepackage{amsmath}
\usepackage{amssymb}
\usepackage{appendix}

\def\be{\begin{equation}}
\def\ee{\end{equation}}
\def\ba{\begin{eqnarray}}
\def\ea{\end{eqnarray}}
\def\go{\mathrel{\raise.3ex\hbox{$>$}\mkern-14mu
             \lower0.6ex\hbox{$\sim$}}}
\def\lo{\mathrel{\raise.3ex\hbox{$<$}\mkern-14mu
             \lower0.6ex\hbox{$\sim$}}}

\newcommand{\hatl}{\hat{\mbox{\boldmath $l$}}}
\newcommand{\out}{{\rm out}}
\newcommand{\rmin}{{\rm in}}
\def\cT{{\cal T}}
\def\bcT{{\mbox{\boldmath ${\cal T}$}}}
\def\bOmega{{\mbox{\boldmath $\Omega$}}}

\begin{document}

\title
{Assembly of Protoplanetary Disks and Inclinations of Circumbinary Planets}
\author{Francois Foucart$^1$
and Dong Lai$^2$
}
\affil{
$^1$Canadian Institute for Theoretical Astrophysics, University of Toronto, Toronto, Ontario M5S 3H8, Canada\\
$^2$Department of Astronomy, Cornell University, Ithaca, NY 14853, USA}

\begin{abstract}
The {\it Kepler} satellite has discovered a number of transiting
planets around close binary stars.  These circumbinary systems have
highly aligned planetary and binary orbits.  In this paper, we explore
how the mutual inclination between the planetary and binary orbits may
reflect the physical conditions of the assembly of protoplanetary
disks and the interaction between protostellar binaries and
circumbinary disks.  Given the turbulent nature of star-forming
molecular clouds, it is possible that the gas falling onto the outer
region of a circumbinary disk and the central protostellar binary have
different axes of rotation.  Thus, the newly assembled
circumbinary disk can be misaligned with respect to the
binary. However, the gravitational torque from the binary produces
a warp and twist in the disk, and the back-reaction torque tends to
align the disk and the binary orbital plane.  We present a new,
analytic calculation of this alignment torque, and show that the
binary-disk inclination angle can be reduced appreciably after the
binary accretes a few percent of its mass from the disk. Our
calculation suggests that in the absence of other disturbances,
circumbinary disks and planets around close (sub-AU) stellar binaries,
for which mass accretion onto the proto-binary is very likely to have occurred,
are expected to be highly aligned with the binary orbits, while disks and planets
around wide binaries can be misaligned. Measurements of the mutual inclinations 
of circumbinary planetary systems can provide a clue to the birth environments of such systems.
\end{abstract}

\section{Introduction}

The extremely precise photometry and nearly continuous observations
provided by the {\it Kepler} satellite have led to the discovery of a
number of transiting planetary systems around stellar binaries.  
At the time of this writing, six such circumbinary systems are known,
including Kepler-16 (with stellar binary period of 41 days and planet
orbital period of 229 days; Doyle et al.~2011), Kepler-34 (28~d,~289~d),
Kepler-35 (21~d,~131~d; Welsh et al.~2012), Kepler-38 (19~d,~106~d;
Orosz et al.~2012a), Kepler-47 (stellar binary orbit 7.45~d, 
with two planets of periods 49.5~d and 303.2~d; Orosz et al.~2012b),
and KIC~4862625 (20~d, 138~d; Schwamb et al.~2012, Kostov et al.~2012).
The stars in these systems have masses of order of the mass of the sun
or smaller, and the planets have radii ranging from 3 earth radii (Kepler-47b) to
0.76 Jupiter radii (Kepler-34b).

By virtue of their detection methods, all the Kepler circumbinary
systems have highly aligned planetary and stellar orbits, with the
mutual orbital inclinations constrained between $\Theta \sim 0.2^\circ$
(for Kepler-38b) and $\Theta \lo 2^\circ$ (Kepler-34b and Kepler-35b).  In
Kepler-16, measurement of the Rossiter-McLaughlin effect further
indicates that the spin of the primary is aligned with the orbital angular
momentum of the binary (Winn et al.~2011).
A natural question arises: do misaligned ($\Theta\go 5^\circ$)
circumbinary planetary systems exist? If so, under what conditions 
can they form?


One might expect that circumbinary systems naturally form with highly
aligned orbits, since the associated orbital angular momenta originate
from the protostellar cores. However, several lines of evidence suggest that
misaligned configurations may be present in some systems:

(i) Solar-type main-sequence binaries with large separations ($\go 40$~AU)
often have a rotation axis misaligned relative to the orbital angular momentum
(Hale 1994). Misalignments are also observed in some short-period binaries, 
such as DI Hercules (with orbital period of 10 days; Albrecht et al. 2009;
see also Albrecht et al.~2011; Konopacky et al.~2012; Triaud et al.~2012).

(ii) Some binary young stellar objects (YSOs) are observed to
contain circumstellar disks that are misaligned with the 
binary orbital plane (e.g., Stapelfeldt et al.~1998).
Also, several unresolved YSOs or pre-main sequence binaries have jets 
along different directions, again suggesting misaligned disks (e.g., 
Davis, Mundt \& Eisl\"offel 1994; Roccatagliata et al.~2011).

(iii) Imaging of circumbinary debris disks shows that the
disk plane and the binary orbital plane are aligned for some systems
(such as $\alpha$ CrB, $\beta$ Tri and HD 98800), and misaligned for
others (such as 99 Herculis, with mutual inclination $\go 30^\circ$;
see Kennedy et al.~2012a,b). Also, 
the pre-main sequence binary KH~15D is surrounded by a precessing
circumbinary disk inclined with respect to the binary plane by
$10^\circ$-$20^\circ$ (e.g., Winn et al.~2004; Chiang \& Murray-Clay 2004; 
Capelo et al.~2012), 
and the FS Tauri circumbinary disk appears to be 
misaligned with the circumstellar disk (Hioki et al.~2011).


While the aforementioned ``misalignments'' may have various origins
(e.g., dynamical interactions in few body systems), in this
paper we focus on the possible existence of warped, misaligned disks
around proto-stellar binaries. We consider scenarios for the assembly
of circumbinary disks in the context of binary star formation (Section
2). These scenarios suggest that circumbinary disks may form with 
misaligned orientations with respect to the binary.
We then study the mutual gravitational interaction between the
misaligned disk and the binary (Section 3) and the long-term evolution
of the binary-disk systems (Section 4). We discuss our results in
Section 5 and conclude in Section 6.

\section{Formation of Binaries and Circumbinary disks: Scenarios}

Binary stars are thought to form by fragmentation inside the
collapsing cores/clumps of molecular clouds, either due to turbulent
fluctuation in the core (``turbulent fragmentation''; e.g., Goodwin et
al.~2007; Offner et al.~2010) or due to gravitational instability in
the resulting disk (``disk fragmentation''; e.g., Adams et al.~1989;
Kratter et al.~2008). In the turbulent fragmentation scenario, the
binaries form earlier and have initial separations of order 1000~AU.
Disk fragmentation also leads to binaries with large initial separations
($\sim 100$~AU). In both cases, continued mass accretion and inward
migration, either due to binary-disk interactions (e.g., Artymowicz
\& Lubow 1996) or dynamical interactions in few-body systems, 
are needed in order to produce close (sub-AU)
binaries. Planet formation can take place in the circumbinary disk
during or after the binary orbital decay.

In the simplest picture, the proto-binary and circumbinary disk rotate
in the same direction. However, molecular clouds and their collapsing
cores are turbulent (see McKee \& Ostriker 2007; Klessen 2011).  It
is natural that the condensing and accreting cores contain gas
which rotates around different directions. Even if the cores are not
turbulent, tidal torques between neighboring cores in a crowded star
formation region can change the rotation direction of the outer regions
of the condensing/accreting cores. Thus the gas that falls onto the central
protostellar core and assembles onto the disk at different times may
rotate in different directions. Such ``chaotic'' star formation has
been seen in some numerical simulations (Bate et al.~2003).  In this
scenario, it is reasonable to expect a rapidly rotating central
proto-stellar core which fragments into a binary, surrounded by a 
misaligned circumbinary disk which forms as a result of continued gas accretion.



The mutual gravitational interaction between a proto-binary
and the circumbinary disk leads to secular evolution of the
relative inclination between the disk and the binary plane.
In most cases, this interaction, combined with continued mass accretion,
tends to reduce the misalignment. We will address these issues
in the next two sections. Note that previous works have
focused on warped {\it circumstellar} disks inclined relative to the
binary (e.g., Papaloizou \& Terquem 1995; Bate et al.~2000;
Lubow \& Ogilvie 2000). The warped/twisted circumbinary disks studied
below have qualitatively different behaviours.




\section{Warped Circumbinary disks}
\label{sec:analytic}

\subsection{Disk-Binary Interaction}

Consider a circumbinary disk surrounding a stellar
binary. The two stars have masses $M_1$ and $M_2$, and are assumed to
have a circular orbit of semi-major axis $a$.  The circumbinary
disk has surface density $\Sigma(r)$, and extends from $r_{\rm in}$
to $r_{\rm out}(\gg r_{\rm in})$. The inner disk is truncated by the 
tidal torque from the binary, and typically $r_{\rm in}\sim 2a$
(Artymowicz \& Lubow 1994; MacFadyen \& Milosavljevic 2008). 
The orientation of the disk at radius $r$ (from the center of mass of the
binary) is specified by the unit normal vector $\hatl (r)$.
Averaging over the binary orbital period and the disk azimuthal direction,
the binary imposes a torque per unit area on the disk element at radius $r$ given,
to leading order in $a/r$, by 
\be
{\bf T}_{\rm b} = -\frac{3}{4} \frac{G M_t\eta\Sigma a^2}{r^3} 
\,({\hatl}_b\cdot\hatl) ({\hatl}_b \times \hatl),
\label{eq:torque}\ee
where $M_t=M_1+M_2$ is the total mass, $\eta=M_1M_2/M_t^2$ 
the symmetric mass ratio of the binary, and ${\hatl}_b$ is 
the unit vector along the orbital angular
momentum of the binary.
\footnote{
A similar calculation, but for the tidal torques imposed on a circumstellar
disk by a binary companion, can be found in Appendix B of Ogilvie \& Dubus (2001).
For circumbinary disks, the only differences are that 
we expand the gravitational potential of the stars to first order in $a/r$ instead of
$r/a$, and consider the motion of both stars around the center of mass of the system
instead of the motion of the companion relative to the primary star.
}
Under the influence of this torque, the 
angular momentum of an isolated disk element would precess at 
the frequency $-\Omega_p\cos\beta$, where 
$\beta$ is the angle between $\hatl_b$ and $\hatl$, and 
\be
\Omega_p(r) \simeq \frac{3\eta}{4} \frac{a^2}{r^2}\, \Omega(r),
\ee
with $\Omega(r)\simeq \Omega_K=(GM_t/r^3)^{1/2}$ the disk rotation rate.
Since $\Omega_p$ depends on $r$, the differential precession 
can lead to the warping (change with $r$ of the angle between $\hatl$ and $\hatl_b$) 
and twisting (change of $\hatl$ orthogonal to the $\hatl-\hatl_b$ plane) of the disk.


\subsection{Dynamical Warp Equations for Low-Viscosity disks}

Theoretical studies of warped disks (Papaloizou \& Pringle 1983;
Papaloizou \& Lin 1995)
have shown that there are two dynamical regimes for the linear propagation 
of warps in an accretion disk. For high viscosity Keplerian disks with $\alpha\go
\delta\equiv H/r$ (where $H$ is the half-thickness of the disk, 
and $\alpha$ is the
Shakura-Sunyaev parameter such that the viscosity is $\nu=\alpha
H^2\Omega$), the warp satisfies a diffusion-type equation with
diffusion coefficient $\nu_2=\nu/(2\alpha^2)$. For low-viscosity
disks ($\alpha\lo \delta$), 
on the other hand, the warp propagates as bending waves 
at about half the sound speed, $c_s/2$.
Protoplanetary disks with $\alpha\sim 10^{-4}$-$10^{-2}$ likely
satisfy $\alpha\lo \delta$ (e.g., Terquem 2008; Bai \& Stone 2011).
For such disks, the warp evolution equations governing
long-wavelength bending waves in the linear regime 
($|\partial\hatl/\partial\ln r|\ll 1$) are given by
(Lubow \& Ogilvie 2000; see also Lubow et al.~2002, Ogilvie 2006)
\ba
\label{eq:dtlvl}
&&\Sigma r^2 \Omega \frac{\partial \hatl}{\partial t} = \frac{1}{r}
\frac{\partial {\bf G}}{\partial r} + {\bf T}_{\rm b}, \\
\label{eq:dtG}
&&\frac{\partial {\bf G}}{\partial t} = \left(\!\frac{\Omega^2 -
  \Omega_r^2}{2\Omega}\!\right) \hatl \times {\bf G} - \alpha \Omega
{\bf G} + \frac{\Sigma H^2 \Omega_z^2 r^3 \Omega}{4} \frac{\partial
  \hatl}{\partial r},
\ea
where $\Omega_r$ and $\Omega_z$ are the radial epicyclic frequency and
the vertical oscillation frequency, ${\bf G}$ is the internal torque
of the disk, and the surface density $\Sigma(r)$ is taken to be the same as that
of the unwarped disk.  These equations are only valid for $\alpha \lo
\delta \ll 1$, $|\Omega_r^2-\Omega^2|\lo \Omega^2\delta$ and
$|\Omega_z^2-\Omega^2|\lo \Omega^2\delta$.
For circumbinary disks considered here, the rotation rate differs from
the Keplerian rate by an amount
$|\Omega^2-\Omega_K^2|/\Omega_K^2 = {\cal O}\left(\eta a^2/r^2
\right)+{\cal O}\left(\delta^2\right)$, and similarly for $\Omega_r$ and
$\Omega_z$ (see Eqs~(\ref{eq:OmNK})-(\ref{eq:OmrNK}) below). So the validity of equations~(\ref{eq:dtlvl})-(\ref{eq:dtG}) 
requires $\eta a^2/r^2\lo \delta$, a condition that is generally satisfied.

In the absence of the external torque (${\bf T}_b=0$),
equations~(\ref{eq:dtlvl})-(\ref{eq:dtG}) admit wave solutions.  
If we define a Cartesian coordinate system so that $\hat l_z \simeq 1$ and
$|\hat l_{x,y}|\ll 1$, then 
a local (WKB) bending wave with $\hatl_{x,y}, {\bf G}
\propto e^{ikr-i\omega t}$ has a phase velocity 
$\omega/k\simeq \pm c_s/2=\pm H\Omega/2$ (assuming $\omega\ll \Omega
\simeq \Omega_r\simeq \Omega_z$).

\subsection{Steady-State Warp and Twist of Circumbinary disks}

We now consider a circumbinary disk whose rotation axis at the outer
radius ($r_{\rm out}$), $\hatl_{\rm out}=\hatl(r_{\rm out})$, is
inclined relative to the binary direction $\hatl_b$ by a finite angle,
$\beta(r_{\rm out})\equiv\Theta$.
This corresponds to the situation where the outer disk
region is fed by gas rotating around the axis $\hatl_{\rm out}$.
In the Cartesian coordinate system with the $z$-axis along $\hatl_b$,
the disk direction $\hatl(r)$ can be written as 
\be
\hatl(r)=(\sin\beta\cos\gamma,\sin\beta\sin\gamma,\cos\beta),
\ee
where $\beta(r)$ is the warp angle and $\gamma(r)$ is the twist angle.
At $r=r_{\rm out}$, we have $\hatl_{\rm out}=(\sin\Theta,0,
\cos\Theta)$ without loss of generality. 
A steady-state warp/twist is reached after a few
bending wave propagation times across the whole disk. The steady-state
warp/twist profile can be obtained by numerically integrating
Eqs.~(\ref{eq:dtlvl})-(\ref{eq:dtG}) and setting $\partial/\partial t
=0$.  Figure 1 depicts selected numerical results.

\begin{figure}
\includegraphics[width=8.3cm]{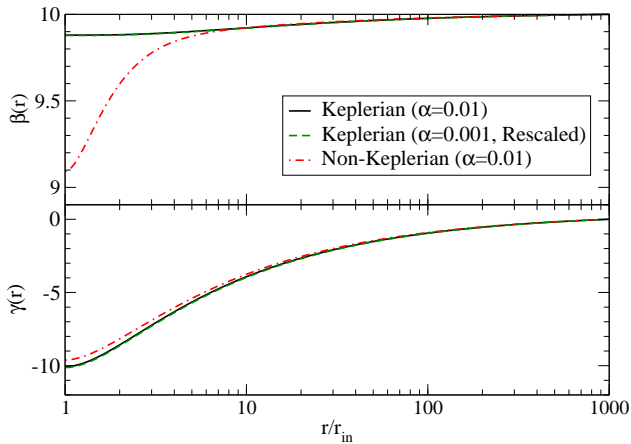}
\caption{Steady-state warp ({\it top panel}) and twist 
({\it bottom panel}) profiles, in degrees, of a
circumbinary disk for which the
  outer disk is misaligned by $10^\circ$ with respect to the angular
  momentum axis of the binary. Both stars in the binary have the same
  mass, and the disk parameters are $p=0.5$ [See Eq.~(\ref{eq:defp})], $\delta=0.1$, 
$r_{\rm in}=2a$, 
and $\alpha=0.01$ or $\alpha=0.001$. The profiles for
  $\alpha=0.001$ are rescaled to show the approximate scaling of the
  warp and twist with $\alpha$ (i.e. the warp is multiplied by 100 and the twist by 10).
  Note that the rescaled $\alpha=0.001$ profiles nearly conincide with the $\alpha=0.01$ Keplerian profiles.
  For $\alpha=0.01$, we show both the Keplerian profile, and results including
  the leading order non-Keplerian correction. The non-Keplerian term significantly
  increases the warp of the disk, but has only a small effect on its twist.}
\label{fig:Profile}
\end{figure}

Physically, the steady-state warp/twist profile is determined by
balancing the internal viscous torque ${\bf G}$ of the disk and the
external torque ${\bf T}_b$.  The viscous damping time of the disk
warp [associated with the viscosity $\nu_2=\nu/(2\alpha^2)$] is
\be
t_{\rm v2}={r^2\over \nu_2}={2\alpha^2 r^2\over \nu}=
{2\alpha\over \delta^2\Omega}.
\ee
A critical warp radius $r_{\rm warp}$ is obtained by equating $t_{\rm v2}$ to 
the precession time $\Omega_p^{-1}$ of an isolated disk element, i.e.
$t_{\rm v2}\Omega_p=1$ at $r=r_{\rm warp}$ where
\be
t_{\rm v2}\Omega_p={3\alpha\eta\over 2}\left({a\over r\delta}
\right)^2.
\label{eq:tvis0}\ee
This gives
\be
r_{\rm warp} \approx a\left({3\alpha\eta\over 2\delta^2}\right)^{1/2}.
\ee
For $r_{\rm warp} \gg r_{\rm in}$, we expect that, in steady state,
the disk well inside $r_{\rm warp}$ be
aligned with the binary $\hatl_b$, while the disk well outside
$r_{\rm warp}$ be aligned with $\hatl_{\rm out}$. However, if the inner disk
radius $r_{\rm in}$ is larger than $r_{\rm warp}$, or $(t_{\rm v2}
\Omega_p)_{\rm in}\lo 1$ (the subscript ``in'' means that the
quantity is evaluated at $r=r_{\rm in}$), then the
whole disk can be approximately aligned with $\hatl_{\rm out}$, with very
small warp between the inner and outer edge of the disk. For standard disk parameters (e.g.,
$\eta\sim 1/4,\,\alpha\sim 10^{-2},\,\delta\sim 0.1$, $r_{\rm in}\sim
2a$), the inequaility $(t_{\rm v2}\Omega_p)_{\rm in}\lo 1$ or
$r_{\rm warp}\lo r_{\rm in}$ is well satisfied.

Equation~(\ref{eq:dtG}) shows that, to first order, changes in the orientation $\hatl(r)$ of the disk are due
to the combination of a term proportional to the internal stress $\bf{G}$ (which
modifies the twist $\gamma$ of the disk), and a term proportional to $\bf{G}\times \hatl$ (which 
causes variations of the warp $\beta$). The second term only exists for non-Keplerian disks,
while the first is only slightly modified by deviations from a Keplerian
profile. We treat these two effects separately by first considering purely Keplerian
disk profiles, to obtain a good approximation for the twist $\gamma(r)$ in the disk,
and then including non-Keplerian effects, which are generally the main source of the warp $\beta(r)$.

For Keplerian disks ($\Omega_r=\Omega_z=\Omega$) and small warps,
we can obtain approximate, analytic expressions for the
disk warp and twist (see Foucart \& Lai 2011 for a similar 
calculation of magnetically driven warped disks).
For concreteness, we consider disk models with constant $\alpha$, and
assume that the surface density and the 
dimensionless disk thickness have the power-law profiles
\be
\Sigma\propto r^{-p},\quad \delta={H\over r}\propto r^{(2p-1)/4},
\label{eq:defp}
\ee
so that $\dot M\sim \nu\Sigma=\alpha H^2\Omega\Sigma=$constant
\footnote{
In practice, these scalings are unlikely to be valid for the entire radial extent of
the disk. As the warp and twist of the disk are mostly due to the torque
acting at small radii (${\bf T}_b \propto r^{-3-p}$), we are only
concerned about the approximate value of $p$ for $r$ close to $r_{\rm in}$.
}.
Equations (\ref{eq:dtlvl})-(\ref{eq:dtG}) then reduce to
\be
\frac{\partial}{\partial r} \left[\left(\frac{r}{r_{\rm in}}\right)^{\!3/2}
\!\frac{\partial}{\partial r} \hatl \right]
= -\frac{4\alpha r {\bf T}_{\rm b}}{(\delta^2r^5\Sigma \Omega^2)_{\rm in}}.
\label{eq:dldr}\ee
We adopt the zero-torque boundary condition
$\partial\hatl/\partial r=0$ at the inner disk radius $r=r_{\rm in}$.
Since ${\bf T}_{\rm b}$ falls off rapidly with $r$ as ${\bf T}_b \propto
r^{-3-p}$, we can integrate Eq.~(\ref{eq:dldr}) approximately to obtain:
\be
\label{eq:lindl}
\left(\frac{r}{r_{\rm in}}\right)^{\!\!3/2}\!\frac{\partial}{\partial r} 
\hatl \simeq \frac{4\alpha}{1+p}\left(\frac{{\bf T}_{\rm b}}
{\delta^2r^3 \Sigma \Omega^2}\right)_{\rm in}\!
\left[\left(\frac{r_{\rm in}}{r}\right)^{\!1+p}-1\right].
\ee
Using the outer boundary condition $\hatl(r_{\rm out})=\hatl_{\rm out}$,
we find
\ba
&&\hatl(r) - \hatl_{\rm out}\simeq 
\left(\frac{8\alpha}{1+p}\right)
{{\bf T}_b(r)\over (\delta^2 r^2\Sigma\Omega^2)_{\rm in}}
\left({r\over r_{\rm in}}\right)^{\!3/2}F_p(r)\nonumber\\
&& ~~=-{4\,t_{\rm v2}\Omega_p\over (1+p)} F_p(r)
(\hatl_b\cdot\hatl)(\hatl_b\times\hatl),
\label{eq:lml}\ea
where 
\be
F_p(r)=\left(\!{r\over r_{\rm in}}\!\right)^{\!\!p+1}\!\!
-{1\over 2p+3},
\ee
and $t_{\rm v2}\Omega_p$ is given by Eq.~(\ref{eq:tvis0}).
In deriving Eq.~(\ref{eq:lml}), we have assumed that the warp and twist 
of the disk are small compared to its inclination relative to the binary
axis, i.e., 
$|\hatl(r) - \hatl_{\rm out}| \ll \sin{\beta}$, or 
$(t_{\rm v2}\Omega_p)_{\rm in}\ll 1$. The total change in $\hatl$ across the disk is then
\be
\hatl_{\rm in} - \hatl_{\rm out}
\simeq -{8(t_{\rm v2}\Omega_p)_{\rm in}\over 2p+3}
(\hatl_b\cdot\hatl_{\rm out})(\hatl_b\times\hatl_{\rm out}).
\ee
The net twist angle across the disk,
$\Delta\gamma_{\rm twist}\equiv \gamma_{\rm in}-\gamma_{\rm out}$, is
\ba
\Delta\gamma_{\rm twist} &\simeq & -{8(t_{\rm v2}\Omega_p)_{\rm in}
\over (2p+3)}\cos\Theta\nonumber\\
&=& -{12\over (2p+3)}\,
\left({\alpha\eta\over \delta_{\rm in}^2}\right)\!
\left({a\over r_{\rm in}}\right)^2\cos\Theta.
\label{eq:deltatwist}
\ea

For Keplerian disks, the warping is only a second-order effect: to first order
$\hatl_{\rm in} - \hatl_{\rm out}$ is perpendicular to $\hatl_{\rm out}$,
and the disk is only twisted. The torque acting on the inner disk
does however have a component in the $\hatl_b-\hatl_{\rm out}$
plane, due to the small difference in orientation
between $\hatl_{\rm in}$ and
$\hatl_{\rm out}$. The net warp angle, $\Delta\beta_{\rm warp}\equiv
\beta_{\rm in}-\beta (r_\out)=\beta_{\rm in}-\Theta$, is given by
\ba
\Delta\beta_{\rm warp} &\simeq &
-\left[{4 (t_{\rm v2}\Omega_p)_{\rm in}\over 2p+3}\right]^2
\sin(2\Theta)
\nonumber\\
&=&-\left[{6\over 2p+3}\,\left({\alpha\eta\over \delta_{\rm in}^2}\right)\!
\left({a\over r_{\rm in}}\right)^{\!2}\right]^2\sin(2\Theta).
\label{eq:betaK}
\ea
As noted before, these expressions for $\Delta\gamma_{\rm twist}$ and
$\Delta\beta_{\rm warp}$ are valid only for $|\Delta\gamma_{\rm twist}|\ll 1$,
or $(t_{\rm v2}\Omega_p)_{\rm in}\ll 1$.

Figure 1 shows that the numerically integrated disk profile
agrees with both the analytic amplitudes of the warp and twist and their 
scaling with the viscosity parameter $\alpha$. Thus, 
for standard disk and binary parameters ($\eta=0.25$, 
$\alpha=10^{-3}$-$10^{-2}$, $\delta=0.1$ and $r_{\rm in}\simeq 2a$),
the steady-state Keplerian disk is almost flat, with its orientation determined
by $\hatl_{\rm out}$, i.e., the angular momentum axis
of the gas falling onto the outer disk.

Deviations from a Keplerian disk profile modify the above results,
as differences between the epicyclic and orbital frequency of the disk
are induced by both the finite thickness of the disk and the deviation
of the binary gravitational potential from its point-mass value. To first order in
$\delta^2$ and $\eta (a/r)^2$, we have (assuming small binary-disk inclination
$\Theta$)
\ba
\label{eq:OmNK}
\Omega^2 &\approx& \frac{GM_t}{r^3} \left( 1 + \frac{3\eta a^2}{4 r^2} - C \delta^2\right)\\
\label{eq:OmrNK}
\Omega_r^2 &\approx&  \frac{GM_t}{r^3} \left(1 - \frac{3\eta a^2}{4r^2} -D \delta^2 \right)
\ea
where $C$ and $D$ are constants of order unity which depend on the density/pressure
profile of the disk, and the epicyclic frequency was computed from 
$\Omega_r^2=(2\Omega/r)d(r^2\Omega)/dr$.
We thus have
\be
\frac{\Omega^2 - \Omega_r^2}{\Omega^2} \approx \frac{3\eta a^2}{2 r^2} + (D-C) \delta^2.
\ee
It is worth noting that for $\delta={\rm constant}$, the $\delta^2$ term vanishes
(since $C=D$ in that case).
Including this result in Eq.~(\ref{eq:dtG}) leads to an additional warp
\be
\label{eq:betaNK}
\Delta \beta_{\rm warp}^{\rm NK} \approx - K \frac{\eta}{\delta_{\rm in}^2}
\left(\frac{a}{r_{\rm in}}\right)^2 \sin(2\Theta)
\ee
with
\be
K = \frac{0.9\eta}{2p+7}  \left(\frac{a}{r_{\rm in}}\right)^2 + \kappa \delta_{\rm in}^2
\ee
and $\kappa$ a constant depending on the profile of $\delta(r)$ close to $r_{\rm in}$ ($\kappa=0$
for constant $\delta$, and of order unity for slowly varying $\delta$).
Numerical results for the non-Keplerian steady-state disk profile are shown in Figure 1
for constant thickness $\delta$.
We see that even though the warp remains relatively small
($\Delta \beta_{\rm warp}^{\rm NK} \sim 0.1 \beta$), for $\alpha \lo 0.03$ it will
be larger than the second-order Keplerian warp given by Eq.~(\ref{eq:betaK}). 
As most of the torque on the disk is due to the contributions
at radii $r\sim r_{\rm in}$, this warp also causes a reduction of the Keplerian twist
$\Delta \gamma_{\rm twist}$ [See Eq.~(\ref{eq:deltatwist})]
by a factor of order $(\sin \beta_{\rm in} / \sin \beta_{\rm out})$.

\section{Evolution of the Relative Binary - disk Inclination}

As discussed in Section 3, the binary torque [Eq.~(\ref{eq:torque})]
induces a small warp and twist in the circumbinary disk. For disks satisfying
$\delta\go \alpha$, the steady-state
warp/twist is achieved when transient bending waves either damp out or propagate out
of the disk (depending on their behavior at large radius). Bending waves propagate
at half the sound speed, and will thus reach the outer boundary of the disk over
a timescale
\be
t_{\rm warp}\sim 
\int_{r_{\rm in}}^{r_\out}{dr\over H\Omega/2}
\sim  {2\over \delta_\out\Omega_\out}.
\ee
As for the damping of transient bending waves, it is due to the $\alpha \Omega {\bf G}$
term in Eq.~(\ref{eq:dtG}). Numerical results (Lubow \& Ogilvie 2000) have confirmed
that the damping timescale is 
\be
t_{\rm damp} \sim {1 \over \alpha \Omega_{\rm out}}.
\ee
Both timescales are much shorter than the age of the system
or the gas accretion time
\be
t_{\rm acc}\sim \left({r^2\over\nu}\right)_\out\sim
{1\over \alpha\delta_\out^2\Omega_\out}.
\ee

On a timescale longer than $t_{\rm damp}$, the warped disk exerts a
back-reaction torque on the binary, aligning $\hatl_b$ with the disk
axis (more precisely, with $\hatl_\out$).  To determine this torque,
recall that in the Cartesian coordinate system that we have set up,
$\hatl_b=(0,0,1)$ and $\hatl_\out=(\sin\Theta,0,\cos\Theta)$.
In the small-warp approximation, $\hatl(r)\simeq
(\sin\Theta + \Delta{\hat l}_x,\Delta{\hat l}_y,\cos\Theta)$,
where $\Delta{\hat l}_{x,y}$ are the ($x$,$y$)-components of
$\hatl(r)-\hatl_\out$.  $\Delta{\hat l}_y$ is well approximated by [see Eq.~(\ref{eq:lml})]
\be
\Delta{\hat l}_y\simeq 
-{4\,t_{\rm v2}\Omega_p\over (1+p)} F_p(r)\cos\Theta\sin\Theta,
\ee
and thus $|\Delta{\hat l}_{y}|\ll\sin\Theta$
when $t_{\rm v2}\Omega_p\ll 1$.
$\Delta{\hat l}_x$ is mainly due to non-Keplerian effects [see Eq.~(\ref{eq:betaNK})],
and is generally a small correction ($\sim 10\%$) to $\sin\Theta$.
Thus the torque on the disk element 
[Eq.~(\ref{eq:torque})] is, to leading order for each component,
\be
{\bf T}_b\simeq 
-\frac{3}{4} \frac{G M_t\eta\Sigma a^2}{r^3} \cos\Theta
\left(-\Delta{\hat l}_y,\sin\Theta,0\right).
\ee
The back-reaction torque on the binary is
\be
\bcT=-\int_{r_{\rm in}}^{r_\out}\! 2\pi r{\bf T}_b\,dr
\ee
The $x$-component of $\bcT$ tends to align $\hatl_b$ with $\hatl_\out$:
\be
\cT_x\simeq {72\pi\over (2p+3)(4p+5)}\left({\alpha\eta^2\over\delta_\rmin^2}
\right)a^4\Sigma_\rmin\Omega_\rmin^2\cos^2\!\Theta\sin\Theta.
\label{eq:Tx}\ee
If we define the angular momentum of the inner disk region by
\be
(\Delta J)_{\rm in}\equiv 2\pi (\Sigma r^4\Omega)_{\rm in},
\ee
then $\cT_x$ can be written as
\be
\cT_x={32\over (2p+3)(4p+5)}(\Delta J)_{\rm in}\,
(\Omega_p^2t_{\rm v2})_{\rm in},
\ee
where $t_{\rm v2}\Omega_p=(3\alpha\eta/2)(a/r\delta)^2$ 
[see Eq.~(\ref{eq:tvis0})]. 

The $y$-component of $\bcT$ on the binary is
\be
\cT_y={3\pi\over 2(1+p)}GM_t\eta\, {a^2\Sigma_\rmin\over r_\rmin}
\cos\Theta\sin\Theta.
\ee
This makes the binary axis $\hatl_b$ precess around $\hatl_\out$ at the
rate
\be
\bOmega_{\rm prec}=-{3\pi\over 2(1+p)}\left({\Sigma_\rmin a^3\over 
M_t r_\rmin}\right)\Omega_b\cos\Theta\,\hatl_\out,
\ee
where $\Omega_b=(GM_t/a^3)^{1/2}$ is the orbital frequency of the binary.
Since $\cT_y$ does not induce permanent change of the inclination angle
$\Theta$, we will not consider it further in this paper.
 
It is worth noting that to leading order the back-reaction torque is independent 
of the  non-Keplerian warp computed in Eq.~(\ref{eq:betaNK}), even when that warp is the main
deviation from the flat-disk profile.
Indeed, $\cT_x$ is proportional to the twist $\Delta \hatl_y$ and $\cT_y$ to $\sin \Theta$. The
only effect of the non-Keplerian warp is thus to modify $\bcT$ by 
a factor of order ($\sin \beta_{\rm in}/\sin \beta_{\rm out}$).

Mass accretion from the circumbinary disk onto the binary 
can also contribute to the alignment torque. The accretion streams
from $r_{\rm in}$ will likely land in both stars, probably through
circumstellar disks (e.g., Artymowitz \& Lubow 1994; 
MacFadyen \& Milosavljevic 2008). Given the complexity of the process, we 
parametrize the alignment torque due to accretion as
\be
\cT_{{\rm acc},x}=g\,\dot M (GM_t r_\rmin)^{\!1/2}
\sin\Theta,
\label{eq:Tacc}\ee
where $g$ is a dimensionless number of order unity. In writing 
Eq.~(\ref{eq:Tacc}), we have used the result of Section 3 that 
the steady-state circumbinary disk is only slightly warped 
[$\beta(r_{\rm in})\simeq \Theta$].
Since the mass accretion rate is given by
$\dot M\simeq 3\pi\nu\Sigma=3\pi\alpha
(\delta^2 r^2\Omega\Sigma)_\rmin$, we can rewrite Eq.~(\ref{eq:Tx}) as 
\be
\cT_x\simeq f\,\dot M (GM_t r_\rmin)^{\!1/2}\cos^2\!\Theta
\sin\Theta,
\label{eq:ctx}\ee
where
\be
f={24\over (2p+3)(4p+5)}\,\eta^2\left({a\over\delta_\rmin r_\rmin}\right)^4.
\ee
The total alignment torque on the binary is then
\ba
&&\cT_{\rm align}=\cT_{{\rm acc},x}+\cT_x\nonumber\\
&&\qquad\quad \simeq (g+f\cos^2\!\Theta)
\dot M (GM_t r_\rmin)^{\!1/2}\sin\Theta.
\ea
Assuming that the angular momentum of the binary, 
$L_b=\eta M_t(GM_ta)^{1/2}$, is
much less than that of the disk (and the material falling onto the
disk), the torque $\cT_{\rm align}$ leads to alignment between
$\hatl_b$ and $\hatl_\out$, on the timescale
\be
t_{\rm align}={L_b\sin\Theta\over \cT_{\rm align}}
={\eta M_t\over\dot M}\left({a\over r_\rmin}\right)^{\!\!1/2}\!\!
{1\over (g+f\cos^2\!\Theta)}.
\label{eq:talign}\ee
The secular evolution of $\Theta(t)$ is determined by the equation
\be
{d\Theta\over dt}=-{\sin\Theta\over t_{\rm align}}.
\label{eq:dtheta}\ee
For $\Theta\ll 1$, this can be easily solved: Starting from
the initial angle $\Theta (t_i)$, the inclination evolves according to
\be
\Theta (t)=\Theta (t_i)\exp \left[-{\Delta M\over \eta M_t}
\left({r_\rmin\over a}\right)^{\!\! 1/2}(g+f)\right],
\label{eq:betat}\ee
where $\Delta M$ is the total mass accreted through the disk 
during the time between $t_i$ and $t$.

\section{Discussion}

The calculations presented in Sections 3 and 4 show 
that a circumbinary disk formed with its rotation axis
$\hatl_\out$ (at large distance) inclined with respect to the binary
angular momentum axis $\hatl_b$ will attain a weakly warped/twisted state,
such that the whole disk is nearly aligned with $\hatl_\out$
(see Section 3). However, the interaction torque between the disk
and the binary tends to drive $\hatl_b$ toward alignment 
with $\hatl_\out$. The timescale of this alignment is given by
Eq.~(\ref{eq:talign}), and the relative binary-disk inclination
$\Theta$ evolves according to Eq.~(\ref{eq:betat}).

Note that both the accretion torque and the gravitational torque contribute 
to the alignment. If only the accretion torque were present
(i.e., $g\sim 1$, $f=0$), the alignment timescale would be of the same order as
the mass-doubling time of the binary ($t_{\rm align}
\sim 4\times 10^7$~yr for $M_1=M_2=1~M_\odot$, $r_\rmin\simeq 2a$ and
$\dot M\sim 10^{-8}M_\odot\,{\rm yr}^{-1}$), and a significant fraction of
the binary mass would have to be accreted 
($\Delta M\sim 0.4M_\odot$) in order to achieve an $e$-fold reduction of
$\Theta$ [see Eq.~(\ref{eq:betat})]. However, the gravitational
torque dominates over the accretion torque, since 
the condition $f\gg 1$ can be satisfied for
a wide range of disk/binary parameters (although $\alpha^2f\ll 1$ 
must be satisifed for our equations to be valid). For example,
for $p=3/2$ (the density index for the minimum solar nebula), and
$\eta=1/4$ (equal mass binary), we have
\be
f\simeq 14\left({0.1\over\delta_\rmin}\right)^4
\left({2\,a\over r_\rmin}\right)^4.
\ee
Thus, the alignment timescale is (for $f\gg 1$ and $\cos^2\Theta\simeq 1$)
\be
t_{\rm align}\simeq 2.5\left(\!{\eta M_t\over 0.5M_\odot}\!\right)
\!\left(\!{\dot M\over 10^{-8}M_\odot/{\rm yr}}\!\right)^{\!\!-1}\!\!
\left(\!{\delta_{\rm in}\over 0.1}\!\right)^{\!\!4}\!
\left({r_{\rm in}\over 2a}\right)^{\!3.5}{\rm Myrs}
\label{eq:talign2}\ee
($\eta M_t$ is the reduced mass of the binary).
The amount of mass accretion needed for an $e$-fold reduction of $\Theta$ is 
[see Eq.~(\ref{eq:betat})]
\be
(\Delta M)_e\simeq 0.05\,(\eta M_t)
\left(\!{\delta_{\rm in}\over 0.1}\!\right)^{\!\!4}\!
\left({r_{\rm in}\over 2a}\right)^{\!3.5}.
\label{eq:dme}\ee
Thus, only a small fraction of the binary mass has to be
accreted to achieve significant reduction of $\Theta$.

We comment on two assumptions adopted in our calculations 
of Sections 3-4: 

(i) We have assumed that the binary separation $a$ is
constant. In reality, the binary-disk alignment can take place simultaneously 
as the binary orbit shrinks (due to binary-disk interactions).
However, since the alignment timescale depends only on the ratio
$r_{\rm in}/a$, we expect our result to be largely 
unchanged in such a situation as long as $r_{\rm in}$ keeps track of
$a$ while the orbit decays.  

(ii) We have assumed that there is a constant supply of gas at the
outer disk and the total angular momentum of the disk, $L_{\rm disk}$,
is much larger than that of the binary, $L_b$. If we consider an
isolated circumbinary disk (e.g., when an episode of mass infall from
the turbulent cloud/core onto the central binary occurs) with $L_{\rm
  disk}$ comparable or smaller than $L_b$, then both the binary and
the disk will precess around a common total angular momentum axis
while the mutual inclination $\Theta$ evolves. In this case,
equation (\ref{eq:dtheta}) is replaced by 
\be
{d\Theta\over dt}=-{\cT_{\rm align}\over L_b}-{\cT_{\rm align}\over 
L_{\rm disk}}\cos\Theta=-{\sin\Theta\over t_{\rm align}}
\left(1+{L_b\cos\Theta\over L_{\rm disk}}\right).
\label{eq:dthetadt}
\ee
This equation neglects the accretion torque,
so that $\cT_{\rm align}\simeq \cT_x$. Note that in general,
$\cT_x$ and $t_{\rm align}$ will be modified from the expressions given in
Section 3. But as long as $r_{\rm out}\gg r_{\rm in}$, we expect 
the corrections to be small (Foucart \& Lai, in preparation).
It is also of interest to note that when $\cos\Theta< - L_{\rm disk}/L_b$,
the gravitational torque tends to drive $\Theta$ toward $180^\circ$
(i.e., counter-alignment). However, this criteria is only valid instantaneously:
both $L_{\rm disk}$ and $L_b$ vary in time due to dissipation in the disk
and accretion onto the binary, and there is thus no guarantee that the evolution
of $\Theta$ is monotonic. To determine which initial conditions
lead to counter-alignment, assumptions have to be made regarding the evolution
of $L_{\rm disk}$ and $L_b$. King et al.(2005) showed that if $L_b$ is constant
(negligible accretion onto the binary), counter-alignment will occur if the less
restrictive condition 
$\cos\Theta < - L_{\rm disk}/(2L_b)$ is satisfied --- even though initially $d\Theta/dt < 0$.
\footnote{
King et al.(2005) studied accretion disks around spinning black holes, not circumbinary disks.
However, the mathematical form of the evolution equation
for $\Theta$ is identical to the circumbinary case
(compare e.g. Eq.~(\ref{eq:dthetadt}) of this work with Eq.(18) of King et al.).
}

\section{Conclusions and Implications}

In this paper, we have considered scenarios for the assembly of
proto-planetary disks around newly formed stellar binaries.  The shape
and inclination of the disk relative to the binary will determine the
orbital orientations of the circumbinary planets that are 
formed in the disk. Because of the turbulence in molecular clouds 
and dense cores, inside which protostellar binaries and circumbinary 
disks form, and also
because of the tidal torques between nearby cores, it is possible,
and even likely, that gas falling onto the outer region of the
circumbinary disk rotates along a direction different from the rotation axis of the
binary. Thus in general, the newly assembled circumbinary disk will be
misaligned with respect to the binary. However, the gravitational
torque from the binary produces a warp and twist in the disk,
and the back-reaction torque associated with that twist 
tends to align (under most conditions) the disk and the binary orbital
plane. We have presented new calculations of the interaction between the
warped/twisted disk and the binary, and showed that the disk warp is
small under typical conditions. More importantly, we have
derived new analytic expressions for the binary-disk
alignment torque and the associated timescale
[see Eq.~(\ref{eq:talign2})].
Our results show that the misalignment angle can be
reduced appreciably after the binary accretes a few precent of
its reduced mass [see Eq.~\ref{eq:dme}].

Proto-binaries formed by fragmentation (either turbulent fragmentation
or disk fragmentation; see Section 2) have initial separations much larger
than 1~AU. Significant inward migration must occur to produce 
close (sub-AU) binaries. Since mass accretion necessarily takes place
during disk-driven binary migration, our results then suggest
that close binaries are likely to have aligned circumbinary disks,
while wider binaries {\it can} have misaligned disks. This can be 
tested by future observations.  The circumbinary planetary systems 
discovered by {\it Kepler} (see
Section 1) all contain close (period $\lo 41$~d) binaries. If the
planets form in the late phase of the circumbinary disk (as is likely to
be the case considering the relatively small planet masses
in the {\it Kepler} systems), then the planetary orbits will be highly aligned
with the binary orbits, even if the initial disk has a misaligned
orientation. This is indeed what is observed. 

Of course, given the complexity of the various processes involved, 
one may see some exceptions.
In particular, in this paper we have not considered any dynamical processes
(few body interactions) that may take place after the binary and planet
formation. Such processes can also affect the mutual inclinations
of circumbinary planets.

Observationally, tertiary bodies on misaligned orbits around close,
eclipsing binaries can be detected by searching for periodic eclipse
timing variations.  This has led to the identification of many
binaries with tertiary companions (e.g., Liao \& Qian 2010; Gies et
al.~2012).  A number of post-main-sequence eclipsing binaries have
been claimed to host candidate circumbinary planets, such as HW
Virginis (Lee et al.~2009), HU Aquarii (Qian et al.~2011; Gozdziewski
et al.~2012), NN Serpentis (Beuermann et al.~2010), DP Leonis
(Beuermann et al.~2011) and NY Vir (Qian et al.~2012).  However,
some of these claims are controversial since the proposed planetary
orbits may be dynamically unstable on short timescales (see Horner et
al.~2011,2012). Currently, no misaligned (inclination $\go 5^\circ$)
circumbinary planets have been confirmed around main-sequence binaries.

Overall, our calculations in this paper illustrate that the mutual
inclinations and other orbital characteristics of circumbinary
planetary systems can serve as a diagnostic tool for the assembly and
evolution of protoplanetary disks and the condition of formation of
these planetary systems.


\section*{Acknowledgments}
This work has been supported in part by the NSF grants AST-1008245,
AST-1211061 and the NASA grant NNX12AF85G.


\end{document}